\documentclass[11pt,twoside]{article}
\usepackage{asp2014}

\aspSuppressVolSlug
\resetcounters

\newcommand{\HI}{{H}{\sc I}}

\newcommand{\nhi}{$N_{HI}$}
\newcommand{\kms}{km\,s$^{-1}$}
\newcommand{\cmsq}{cm$^{-2}$}

\begin{document}

\title{Neutral Atomic Hydrogen in the Local Universe}
\author{D.J. Pisano,$^{1,2}$ Fabian Walter,$^3$ and Sne\v{z}ana Stanimirovi\'c$^4$}
\affil{$^1$Dept. of Physics \& Astronomy and Center for Gravitational Waves \& Cosmology, West Virginia University, Morgantown, WV, USA\\
 $^2$Adjunct Astronomer at Green Bank Observatory, Green Bank, WV, USA; \email{djpisano@mail.wvu.edu}}
\affil{$^3$Max Planck Institute f\"ur Astronomie, Heidelberg, Germany; \email{walter@mpia.de}}
\affil{$^4$Dept. of Astronomy, U. Wisconsin-Madison, Madison, WI, USA; \email{sstanimi@astro.wisc.edu}}

\paperauthor{D.J. Pisano}{djpisano@mail.wvu.edu}{}{West Virginia University}{Dept. of Physics & Astronomy}{Morgantown}{WV}{26501}{USA}
\paperauthor{Fabian Walter}{walter@mpia.de}{}{Max Planck Institute f\"ur Astronomie}{}{Heidelberg}{}{}{Germany}
\paperauthor{Sne\v{z}ana Stanimirovi\'c}{sstanimi@astro.wisc.edu}{}{UW-Madison}{Dept. of Astronomy}{Madison}{WI}{53706}{USA}


\section{Introduction}

The Next Generation Very Large Array (ngVLA) will play a crucial role in the studies of 21-cm atomic neutral hydrogen (\HI) emission in and around galaxies in the local Universe.  
The unprecedented sensitivity and resolution of the ngVLA will provide key information on the atomic gas reservoir in the circumgalactic medium (CGM) 
and its flow onto the gas disks of galaxies.  The ngVLA will connect this large scale view of \HI\ to individual star-forming cores to study the relatively unexplored transition 
from atomic to molecular gas in nearby galaxies and ultimately how gas is converted into stars.  

In order to understand how gas flows onto galaxies and is then converted into molecular gas and, eventually, stars, there are three regimes to study:  
\HI\ around galaxies in the circumgalactic medium; \HI\ within nearby galaxies; and \HI\ in the Milky Way.  We will take these
up in turn.  

\section{\HI\ in the Circumgalactic Medium}

The extent, morphology and dynamics of the gas in the very far outskirts of galaxies are still
essentially unexplored observationally. These outer parts of galaxies are the interface between the
inner star-forming disk and the cosmic web, and likely these are also the regions
where galaxies accrete gas, possibly through the ``cold accretion" process \citep[e.g.][]{keres05}.
Measuring the very extended gas distributions in galaxies, and its connection to the cosmic web,
will be a main science driver for the ngVLA.

To date, most of the exploration of the CGM has come through UV absorption line studies.  
The COS-HALOS project \citep{tumlinson13,werk14} has used background quasars to study the 
Lyman$\alpha$ absorption in the halos of low redshift galaxies.  The project has found that \HI\ 
absorption at \nhi$\gtrsim$10$^{14}$\cmsq\ is ubiquitous out to 150 kpc for star-forming galaxies and present in 75\% of
passive galaxies as well \citep{tumlinson13}.  This cool CGM gas represents 25\%-45\% of the total baryon mass
within the virial radius of the galaxy \citep{werk14}\footnote{Our current understanding of the state of the CGM is 
well-summarized in the review by \citet{tumlinson17}}.  Unfortunately, above \nhi$\sim$10$^{16}$\cmsq\ saturation of
absorption lines makes it difficult to get an accurate measure of \nhi; these are the Lyman Limit Systems.  While
below \nhi$\sim$10$^{16-17}$\cmsq, \HI\ absorption is common, particularly in the intergalactic medium, it is impossible
to image in 21-cm \HI\ emission.  Obtaining deep \HI\ emission observations with the ngVLA 
is the only way forward: while Lyman-$\alpha$ absorption observations can reach very low column densities, 
their pencil-beam nature make it extremely difficult to reconstruct the full gas distribution and its kinematics.

\begin{figure}
\includegraphics[width=\textwidth]{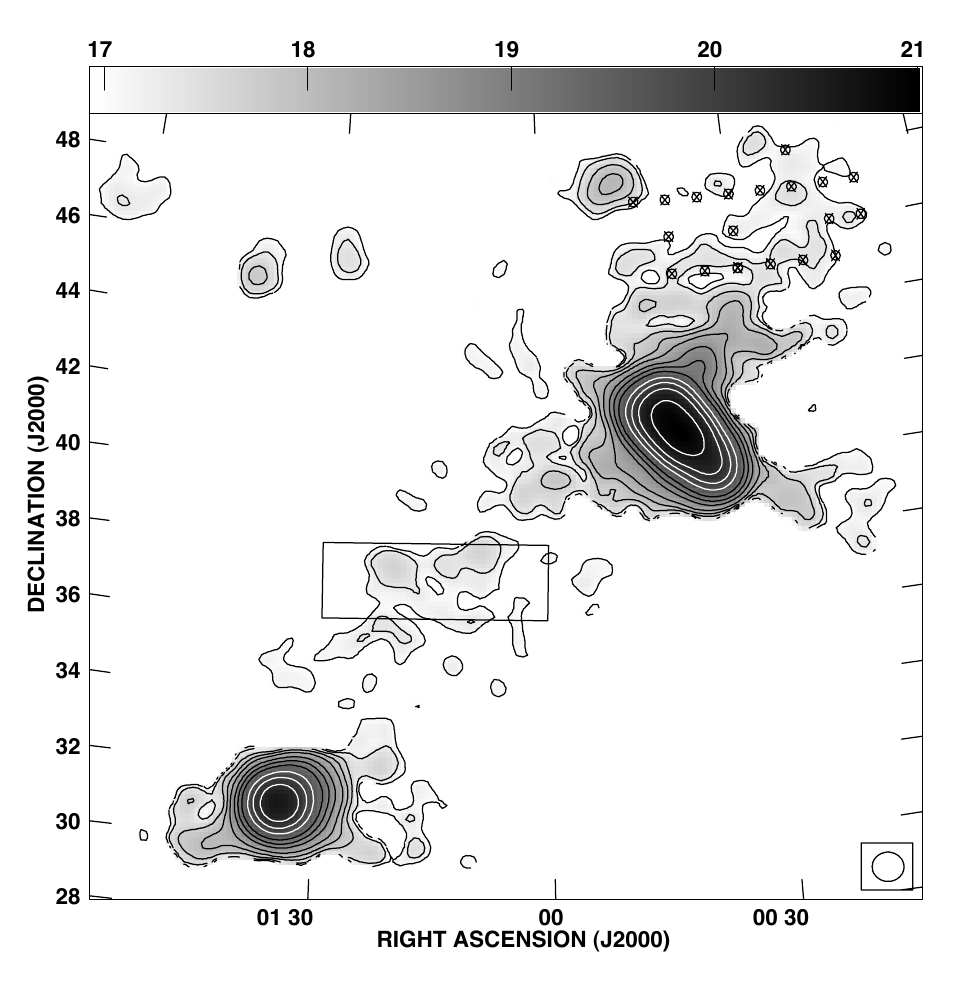}
\includegraphics[width=\textwidth]{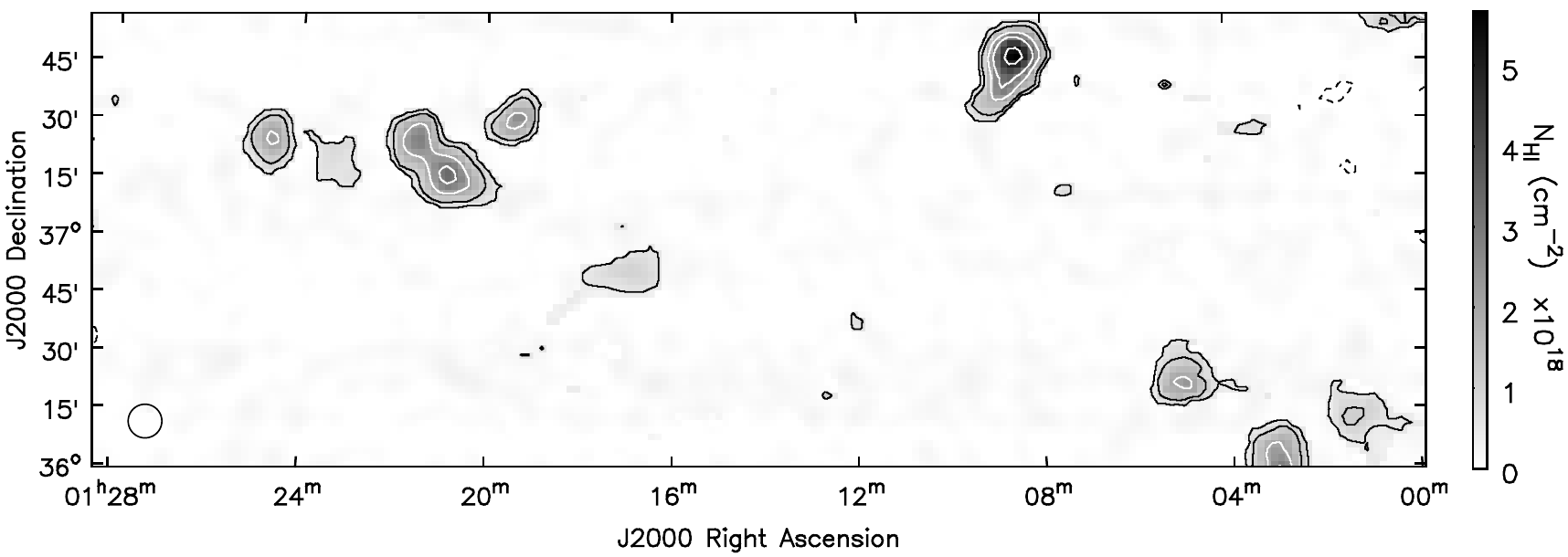}
\caption{Top:  A map of the total \HI\ emission associated with M~33 (lower left) and M~31 (upper right) from \citet{braun04}.  
The contours are at log \nhi$=$17.0, 17.3, 17.7, 18.0, 18.3, 18.7, 19.0, 19.3, 19.7, 20.0, 20.3, and 20.7 [\cmsq].  The beamsize
is shown in the lower right of the panel.  The box shows the region mapped by \citet{wolfe16} with the GBT.  Bottom:  
The GBT \HI\ map from \citet{wolfe16}.  The contours are at -1, 1, 2, 4, 6, and 10 times 5$\times$10$^{17}$\cmsq.  The beam 
size is the circle in the lower left of the image.  Note that the \HI\ structures detected by \citet{braun04} are revealed to be much smaller, 
higher-\nhi\ features by \citet{wolfe16}, illustrating the critical importance of resolution and sensitivity. \label{fig:m31}}
\end{figure}

In galaxy disks the \HI\ surface density is declining as a function of radius, but
observations with current L-band facilities are typically not sensitive enough (reaching column
density sensitivities just below 10$^{20}$\cmsq) to map the \HI\ distribution at larger radii and
correspondingly lower column densities. Does the azimuthally averaged \HI\ surface density profile
continue to decline at the same rate as in the inner disk, or do other effects start to play a role in
determining the shape of the very outer radial \HI\ profile? For example, when \HI\ column densities
decrease, one expectation is that the density of the neutral gas is no longer sufficient for self shielding
and this could cause the majority of the gas to become ionized by the intergalactic
radiation field. Models that include this ionization from the intergalactic radiation field predict that
the steep radial decline of neutral gas surface density will transition, below column densities of
$\sim$10$^{18}$\cmsq, to a more extended, low column density outer disk with a flatter radial profile
\citep[e.g.][]{maloney93,popping09,braun12}, but many of the details are model-dependent.
Gas accretion is not included in these models, and it is currently not known to what degree this
process will affect the outskirts of the gaseous disks of galaxies. Simulations and some tentative
observations indicate the presence of low-column density, kpc-sized gas features near galaxy disks
\citep[e.g.][]{braun04,popping09,wolfe13,wolfe16}.  It is certainly conceivable that these could also 
have a big impact on the \HI\ distribution at very low column densities. In order to detect such features, 
it is important to have both sensitive observations and resolution that is well matched to the size of the 
emitter.  This can be seen in Figure~\ref{fig:m31}.  The ngVLA will have both the sensitivity and resolution
to map this diffuse \HI\ at low \nhi\ for a large number of galaxies.

Besides characterizing the sizes of the gas disks in nearby galaxies, observing the extended \HI\ distribution
will at the same time constrain the numbers of individual \HI\ clouds around the target galaxies.
Based on the relative kinematics and morphologies/locations, one can distinguish between tidal
dwarfs, compact high velocity clouds (HVCs), and tidal material that could be present around the
galaxies under consideration. This will provide a link to the missing galaxy problem, i.e. satellites
with clumps of cool \HI, but no star-formation. Cloud sizes and masses will then provide key
insights into the physical state of gas in the CGM, which can be compared to
cosmological simulations of the structure of the CGM of galaxies \citep[e.g.][]{shen13}. 
Such simulations indicate that the most compelling signature of the cold accretion
phenomenon onto galaxies is a high covering factor $\sim$20-30\% of Lyman limit system absorption in
the CGM on scales out to $\sim$100 kpc (scales that can be easily covered with the ngVLA with small
mosaics).

In addition to this deep ngVLA \HI\ imaging, newly developed techniques are now available to obtain
azimuthally averaged measurements at previously inaccessible radii based on the shifting and
stacking technique described in \citet{ianjamasimanana12}. This method uses an extrapolation
of the (typically flat) rotation curve to create a model velocity field at large radii well beyond that
of the directly detected \HI\ disk. This is then used to shift all spectra in the data cube in velocity by
the amount indicated by the extrapolated velocity field at that position. This ensures that all
(potential) \HI\ signals are lined up at the same reference velocity. Using the same tilted-ring model
that was used for the shifting step, the spectra are then selected in annuli of increasing galaxy
radius and are stacked. The stacked spectrum will have a lower noise value, and if any underlying
HI signal has been lined up through the shift process, the signal-to-noise ratio will have increased
and signatures of otherwise undetected very low column density \HI\ can be detected. Testing this
methodology using THINGS and HALOGAS data has demonstrated that the \HI\ column density
limit can be pushed down by an order of magnitude \citep{ianjamasimanana18}. The
combination of ngVLA's sensitivity, together with the above stacking technique thus will enable
measurements at unprecedented \HI\ columns ($\ll$10$^{17}$\cmsq), that have only been accessible
via Lyman $\alpha$ absorption measurements. In summary, through the combination
of ngVLA observations and new techniques to enhance \HI\ column density sensitivities through
stacking will reach depths in \HI\ column densities that were previously inconceivable. This
measurement will provide the currently completely unexplored connection to the cosmic \HI\ web.

\section{\HI\ in Nearby Galaxies}

Understanding the processes that drive star formation in galaxies is one of the most challenging astrophysical topics, both 
observationally and theoretically.  Of particular interest are studies of how stars form out of the interstellar medium (ISM) on 
scales comparable to the Jeans mass and Jeans length of neutral atomic clouds, and how, in turn, these stars shape the 
structure and physical properties of their ambient ISM.  Of key importance n any such study are high resolution observations 
of the different tracers of the ISM (in particular \HI\ and CO observations) and the stellar population (e.g. broad band and 
H$\alpha$ imaging).  For such studies, resolutions of 5-150 pc are needed to resolve the giant molecular cloud (GMC) complexes 
and the related individual sites of star formation.  It is also on these scales where the feedback of individual supernovae and/or 
massive stars is expected to be most profound. 

Although arcsecond resolutions are routinely achieved using ground--based optical telescopes, it is
observations of similar resolutions at radio wavelengths, in particular the \HI, that are lacking. In
fact, to date, this is the limiting factor of all coherent ISM studies on these small, critical scales. In
the era of ALMA, observations of the molecular gas phase (using CO as a tracer) are now
routinely pushed to arcsecond resolutions. These CO observations are typically limited to the
central few arcminutes of the target galaxies (given by the primary beam size of millimeter
interferometers) and do not recover the atomic gas phase -- they thus cannot be used to
understand global star formation processes in galaxies.  Recently observations of OH emission in the Milky Way have
been shown to trace molecular gas that is unseen in CO emission \citep{allen12, allen15}.  The ngVLA may be able to detect such molecular gas
clouds in other galaxies.  

\begin{figure}
\includegraphics[width=\textwidth]{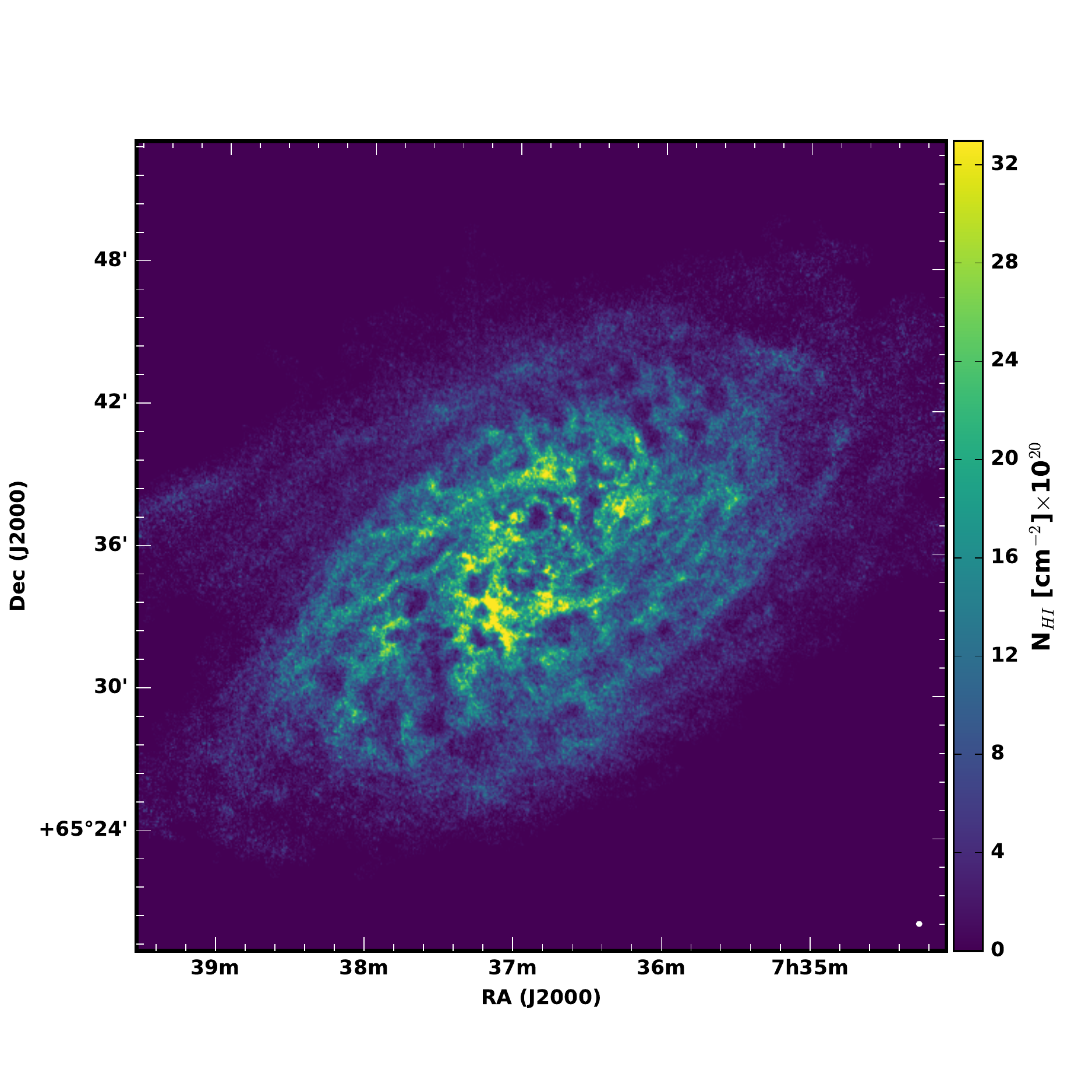}
\caption{A total \HI\ intensity (moment 0) image of NGC~2403 from THINGS \citep{walter08}.  These data were imaged with robust
weighting resulting in a resolution of $\sim$6$\arcsec$.  At the distance of NGC~2403, 3.2 Mpc, this corresponds to a physical resolution
of 93 pc.  The ngVLA will be able to achieve better resolution and sensitivity for larger samples of galaxies in the same amount of time
as the VLA.\label{fig:ngc2403}}
\end{figure}

On the contrary, most nearby galaxies are small enough to fit within the primary beam of existing radio interferometers
(FWHM$_{VLA}\sim$30$\arcmin$) for \HI\ observations and the angular resolution is the limiting factor.  Typically resolutions
of order 15$\arcsec$ are achieved; only the VLA has been pushing these resolutions down to $\sim$6$\arcsec$ for large
samples of galaxies.  The \HI\ Nearby Galaxy Survey \citep[THINGS;][]{walter08} demonstrated that 6$\arcsec$ resolution
can be achieved for a sample of 35 galaxies of any size (i.e. dwarfs to spirals) with the VLA.  Only a couple of galaxies have
been imaged in \HI\ emission at higher resolution, $\sim$1.6$\arcsec$, with the VLA \citep{vanZee06}, but this requires an
extremely large time commitment with the VLA.  The ngVLA will fundamentally change this
situation.  The possibility of mapping nearby galaxies at linear resolutions of 5-150 pc opens the
prospects of exploring processes in the ISM on scales which have hitherto been inaccessible.  Although some
notable exceptions exist, such as: the LMC: \citep{kim03}; the SMC: \citep{staveley-smith97}; 
IC 10: \citep{wilcots98}; and other Local Group dwarfs: \citep{begum06}, but these galaxies are not representative of the galaxy main 
sequence at low redshift.  Figure~\ref{fig:ngc2403} shows a THINGS image of a typical spiral galaxy at $\sim$90 pc resolution; ngVLA will achieve
better resolution and sensitivity for larger samples of galaxies in the same amount of time that the VLA took for THINGS.     

(Sub-)arcsec resolution is commonly reached through massive optical/NIR imaging efforts (ground
and space based, currently with HST, and in the near future, JWST [dust/PAH emission, embedded
SFR tracers]). Resolution-matched \HI\ imaging will enable measurements on the ISM conditions
required for H$_2$ formation \citep{krumholz09,krumholz13}, and the distribution of, e.g., the \HI/H$_2$
ratio will constrain the evolution of molecular clouds. For example, recent sub-pc scale studies of
GMCs in the Milky Way have shown that the \HI\ surface density and the H$_2$ fraction agree well
with the equilibrium model predictions \citep[e.g.][]{lee12}. ngVLA \HI\ observations of Local
Group galaxies will provide crucial tests for the \HI\ surface density and the H$_2$ fraction on pc-scales
in galaxies of varying metallicity and interstellar radiation field required to test both equilibrium
and non-equilibrium models for H$_2$ formation and molecular cloud evolution. Observations at
GMC-scale resolution are also crucial to study turbulence in the ISM. It is at these scales that
energy gets injected into the ISM which then cascades down to smaller scales. We can test if the
\HI\ power spectra indeed follow the predicted power-law scales for turbulence like that found in
the SMC \citep{stanimirovic99,elmegreen01} in a large range of extragalactic environments.

ngVLA GMC-resolution imaging of the \HI\ emission will allow us to study the fine scale structure of
the ISM in galaxies of different types. To date, the only galaxies for which similar linear resolutions
have been achieved have been Local Group dwarfs, notably the LMC/SMC \citep[e.g.][]{kim03}.
These studies have shown how the emission breaks up in filaments and shells, even at the smallest
currently accessible scales. The presence of fractal-like structure in the \HI\ emission is also evident
from recent \HI\ imaging of our Galaxy \citep[in particular the large VLA THOR project]{beuther16}. 
The ngVLA will reach sensitivities and column densities that this work can be replicated in
galaxies across the Hubble sequence, including main sequence galaxies in the local universe. GMC-scale
resolution represents a critical and fundamental scale as it samples the typical sizes of
individual giant molecular and atomic clouds that are the precursors of star formation.
Observations at this resolution will also enable us to directly study the feedback/energy input from
individual supernovae due to massive stars.

A unique feature of the high-resolution \HI\ imaging is that they will, for the first time, enable us to
study non-circular motions in the \HI\ disks of galaxies at scales that are directly relevant for the
$\Lambda$CDM cusp/core controversy. \HI\ observations have the immediate advantage that the kinematics
in the very centers of galaxies can be immediately linked to their outskirts. The large range of
Hubble types that can be studied (including the important dwarf galaxies whose kinematics/DM
content cannot be studied through ALMA imaging using CO) will also enable precise quantification
of the effects of the non-circular motions on the dark matter profile as a function of e.g. disk mass,
depth of the potential, etc. Measuring the strength of shocks and other non-rotation velocity
components at these scales (e.g. across spiral arms/bars) will allow us to derive independent
estimates of the mass in the disk \citep[see e.g. descriptions in][]{weiner01a,weiner01b}, another crucial input
in the derivation of dark matter profiles in galaxies. This information will also provide important
insights in assessing the local stability of gas clouds against gravitational collapse.

GMC-scale resolution data will resolve the giant atomic cloud complexes for the first time in
galaxies beyond the local group and will allow investigations of if and how the peak \HI\ temperature
of the warm \HI\ changes of a function of galaxy type. Of particular interest is the question if peak
\HI\ temperature is different in low-metallicity dwarfs as compared to more metal-rich spirals. Such
studies will directly constrain the physical conditions of the interstellar medium on sub-100 pc
scales. In addition, background continuum sources will be bright enough to perform \HI\ absorption
studies. In this context, ngVLA's 1$\arcsec$ resolution is essential to isolate the absorption from the
surrounding emission. This then opens the possibility to determine directly the \HI\ spin
temperature, something which has largely only been possible in Local Group galaxies \citep[e.g.][]{dickey93,braun97}.

\section{Cloud Scale Imaging of \HI\ in the Milky Way}

The increased sensitivity of the ngVLA will be crucial for addressing major outstanding questions in
ISM physics. The diffuse neutral ISM exists in two flavors: the cold neutral medium (CNM) and the
warm neutral medium (WNM). The CNM is easily studied with the 21-cm line absorption and its
properties have been measured extensively, while only several measurements of the WNM
temperature exist so far. The main reason for this observational paucity is the low optical depth of
the WNM, $\tau\sim$10$^{-3}$. This makes the WNM one of the least understood phases of the ISM. To
constrain theoretical and numerical models of the ISM, temperature distributions over the full
temperature range from $\sim$20 to $\sim$10$^4$ K are essential. Using the upgraded Very Large Array (project
21-SPONGE), \citet{murray14,murray15,murray17} demonstrated that already the VLA can detect \HI\
absorption lines with a peak optical depth $\sim$10$^{-3}$. However, the existing sample size of the current
measurements is of order ~50 directions - way too low to start building proper statistical samples
for a comparison with numerical simulations and understanding of how the CNM-WNM phase mix
vary with interstellar environments.

In addition, the \HI\ excitation processes are still not fully understood. Using a stacking analysis,
\citet{murray14} detected statistically the presence of a widespread WNM population with T$_s$=7200 K. 
This demonstrated that the non-collisional excitation of \HI\ is significant even at high
Galactic latitudes. As Ly$\alpha$ scattering is the most likely candidate for additional excitation of \HI,
the Murray et al. results show that the fraction of Ly$\alpha$ photons, and/or the photon propagation
throughout the ISM, are likely more complicated than what is commonly assumed. For example,
both a theoretical study by \citet{liszt01} and state-of-the-art numerical simulations \citep{kim14} assume 
a uniform flux of Ly$\alpha$ photons throughout the ISM and result is the expected T$_s <$ 4000 K. In 
that regard, the high sensitivity enabled by the ngVLA will allow direct measurements of
WNM spin temperature and its spatial variations in the Milky Way by observing hundreds to
thousands of directions with $\sigma_T <$0.05. By stacking selectively different interstellar environments 
will be probed for the first time.  Finer scale variations may also be possible to measure
by imaging resolved continuum sources to map out variations on sub-parsec scales \citep[e.g.][]{faison98, lazio09, roy12}.

Simultaneously, the ngVLA will be able to map out the raw fuel for H$_2$ formation across major
GMCs in the Milky Way. While it is commonly assumed that the CNM is the key ingredient for
making H$_2$ because of its high density, the observational evidence whether GMC precursors are in
the form of dense \HI\ clouds, diffuse \HI, or smaller molecular clouds, is still lacking \citep{dobbs13}. 
In addition, numerical simulations suggest that GMCs grow via accretion of the
CNM from the envelopes. Recent Green Bank Telescope results, \citep{allen12,allen15}, have shown that
OH is tracing diffuse molecular gas that is not seen via CO measurements.  This may represent an intermediate
stage in forming GMCs.  Constraining where the fuel for H$_2$ formation is in and around GMCs,
what are its physical properties, and what is the accretion rate of this material onto the GMCs, are
essential next steps. Due to sensitivity limitations, almost all \HI\ absorption measurements (e.g. 21-
SPONGE) have focused on random directions, utilizing strong background radio sources. The
ngVLA can provide, for the first time, dense grids of \HI\ absorption spectra in the direction of
selected GMCs. The ngVLA has thus potential to revolutionize our understanding of the ISM
physics, in particular the critical phase transition from the WNM to the CNM to H$_2$ formation.

\section{Limitations of Current Astronomical Instrumentation}

Current radio interferometers lack the combination of resolution and surface brightness sensitivity.  In addition, the larger dishes
of the VLA and WSRT restrict the field of view limiting the region of study.  These all combine to make it near impossible to map
the low-\nhi\ gas within the virial radius of nearby galaxies, while being able to resolve any features.  While single-dish telescopes,
like the GBT, have the surface brightness sensitivity, they lack the resolution to map filamentary structures beyond the Local 
Group.  For high resolution studies of \HI\ in nearby galaxies or in the Milky Way, the sparse nature of the VLA or WSRT mean
that a lot of flux is missing in these data.  

In terms of planned interferometers, MeerKAT is nearing completion and will be well-suited to mapping the diffuse, low-\nhi\ gas
around nearby galaxies; the MHONGOOSE survey \citep{deblok17} will do exactly this for a sample of 30 galaxies.  The IMAGINE survey, 
currently being undertaken with the Australia Telescope Compact Array (ATCA) and the Parkes radio telescope is also studying the diffuse \HI\ around galaxies, 
but both of these surveys will only reach the needed sensitivity over a beam size of $\sim$30$\arcsec$.  Neither survey will have the high 
resolution, $\sim$1$\arcsec$ that will be needed to study the internal structure of galaxies at sufficient detail.  

\section{Unique Capabilities of the ngVLA for \HI\ Science}

The planned technical capabilities of the ngVLA at 1.4 GHz should be matched by the SKA.  The SKA, however, will be located in the southern hemisphere leaving many
nearby galaxies (e.g. M~31) and the outer portions of the Milky Way inaccessible.  Furthermore, current plans are for the SKA to operate in a survey mode.  The ngVLA has
the opportunity to be PI-driven allowing for a wider variety of science to be done.  In addition, for sources that extend over a large area, total power \HI\ measurements 
will be required.  While the ngVLA has a dense core and will be able to recover flux on much larger angular scales than the VLA, short spacing data will still be needed for
some observations.  This can be provided with existing instrumentation using the Green Bank Telescope \citep{frayer17}.  Finally, since ngVLA will be able to operate at higher frequencies than the SKA, much
of the science that requires complementary multi-wavelength data (see below) will require the ngVLA, even if the \HI\ data comes from SKA observations.  

\section{Experimental Layout}

The reference configuration of the ngVLA, Spiral214, can be divided into three parts: the ``core" array, the ``plains" array, and the full array.  For the science described here, only the core array 
(with B$_{max}\sim$1 km) and the plains array (with B$_{max}\sim$30 km), will be useful (ngVLA memo 17).  Some \HI\ absorption studies, such as using background radio galaxies to map small-scale, high-\nhi\ gas,
may use the full array.  Using the reference design sensitivity, the core array will be able to reach $\sigma_{NHI}\sim$10$^{18}$\cmsq\ in 1 hour over a 10 \kms\ channel with a beamsize of $\sim$40$\arcsec$, 
while the plains array will reach $\sigma_{NHI}\sim$6$\times$10$^{20}$\cmsq\ for the same time and channel width over a beam of $\sim$1$\arcsec$.  

For studies of the CGM of nearby galaxies, astronomers will want to use the ngVLA to study \HI\ at \nhi$\sim$10$^{17-18}$\cmsq\ for a $\sim$20 \kms\ linewidth.  To get a 3$\sigma$ detection of 
\HI\ at \nhi$=$10$^{18}$\cmsq\ will require $\sim$30 hours.  To reach \nhi$=$10$^{17}$\cmsq\ would require $\sim$3000 hours at a  resolution of 43$\arcsec$.  If we degrade the resolution to $\sim$1$\arcmin$, 
then we could do this in about 600 hours for a single target.  These time requirements roughly match those of the SKA1-MID telescope \citep[see][for examples]{deblok17}.  Given the planned, semi-random 
distribution of antennas in the core array, we can anticipate that the sensitivity will improve with additional tapering.  As diffuse \HI\ emission could be extended over large angular scales, single-dish, total power
\HI\ data may be needed.  This can easily be provided by the GBT using an array receiver, such as FLAG \citep{roshi18}.  Since the \HI\ filaments expected from models of cold flow accretion are on the scale of 
1--10 kpc across \citep[e.g.]{popping09}, we can probe to \nhi$\lesssim$10$^{17}$\cmsq\ for a small sample of nearby galaxies (D$<$10 Mpc) with relatively modest time requirements ($<$1000 hours) and 
we can study a much larger sample of galaxies out to larger distances at \nhi$=$10$^{18}$\cmsq\ in a similar amount of time.  

For studies of \HI\ within nearby galaxies, we will want the high angular resolution of the plains array ($\theta \sim$1$\arcsec$).  Samples of galaxies at distances of 1~Mpc$<$D$<$30~Mpc are ideal targets--and a wide 
range of different types of galaxies can be studied in this volume.  At these distances, 1$\arcsec$~resolution corresponds to spatial resolutions of 5-150 pc.  In 1 hour, for a 10 \kms\ channel width, the 
ngVLA will yield $\sigma_{NHI}\sim$6$\times$10$^{20}$\cmsq.  Even this sensitivity will allow astronomers to study the high-\nhi\ associated with the gas transitioning between the atomic and molecular phases.  In a 
couple of days of observing we could then redo a survey like THINGS at much higher physical resolution and sensitivity.

In addition to direct observations of \HI\ in nearby galaxies and the Milky Way, we would want to get commensal observations of OH, radio recombination lines, and radio continuum with full polarization.  These data
will help map out the full transition of gas from the atomic to the molecular to the ionized phase.  

\section{Complementarity}

First and foremost, the ngVLA study of \HI\ within nearby galaxies, including the Milky Way, will be complemented by the ngVLA's studies at other wavelengths of other molecules in these systems, such as 
CO, HCN, NH$_3$, etc., as well as the radio continuum associated with star formation and heated dust.  This will yield one of the most detailed pictures of how gas is accreted onto galaxies, transitions into 
molecular clouds, and then turns into stars.  Complementing the data from ngVLA will be a broad range of facilities.   In the infrared, from the near IR through the far IR, will be JWST, Euclid, SPICA, WFIRST, and 
ALMA.  These telescopes will be able to study warm dust, important cooling lines in the ISM (such as [CII]), protostars and obscured star formation in general.  

In the study of the CGM around galaxies, new X-ray facilities such as eROSITA and Athena should allow for the study of the hot halos of galaxies, either directly or via stacking.  Unfortunately, at present
there are no planned UV telescopes that could probe the CGM and IGM in absorption, tracing the WHIM with better sensitivity and towards a denser grid of background sources around many galaxies.  Such a mission
would allow astronomers to both map the hot gas in the X-rays, the warm-hot ionized gas in the CGM using UV absorption lines, and then diffuse, warm neutral hydrogen with the ngVLA.  This would give us a complete
census of the gas available to accrete onto galaxies and fuel future star formation in the local universe.


\acknowledgements The authors thank Nickolas Pingel for his assistance in making figures.  

\bibliography{references.bib}{}  


\end{document}